\documentclass[twocolumn,letter]{jpsj2} %% for letters
\usepackage{amsmath}
\usepackage{graphicx}
\usepackage{dcolumn}
\usepackage{bm}
\newcommand{\llangle}{\langle\kern -.23em \langle}
\newcommand{\rrangle}{\rangle\kern -.23em \rangle}
\renewcommand{\vec}[1]{\boldsymbol{#1}}

\renewcommand{\vec}[1]{\boldsymbol{#1}}
\newcommand{\sgn}{\mbox{sgn}}

\title{Parallel Dynamics of Continuous Hopfield Model Revisited}
\author{
  Kazushi \textsc{Mimura}
  \thanks{E-mail address: {\tt mimura@hiroshima-cu.ac.jp}}
}
\inst{
  Graduate School of Information Sciences, Hiroshima City University, Hiroshima 731-3194, Japan
}
\abst{
We have applied the generating functional analysis (GFA) to a continuous Hopfield model. 
We have also confirmed that the GFA predictions in some typical cases exhibit good consistency with computer simulation results. 
When a retarded self-interaction term is omitted, 
the GFA result becomes identical to that obtained using the statistical neurodynamics as well as the case of the sequential binary Hopfield model. 
}
\kword{generating functional analysis, statistical neurodynamics, exact solution, continuous Hopfield model, zero-temperature Langevin dynamics}

\begin{document}
\maketitle

%~~~~~~~~~~~~~~~~~~~~~~~~~~~~~~~~~~~~~~~~~~~~~~~~~~~~~~~~~~~~~~~~~~~~~~~~~~~~~~~~~~~~~~~~~~~~

In the field of information theory, various important inference problems are described by continuous nonlinear systems which include disorder. 
Some examples are code-division multiple-access (CDMA) \cite{Tanaka2001,Tanaka2005,Kabashima2003,Mimura2005a}, 
error-correcting codes \cite{Sourlas1989,Kabashima1998,Kabashima2000,Kabashima2004,Nishimori1999,Montanari2000}, 
lossy compression \cite{Murayama2003,Murayama2004,Hosaka2002,Hosaka2005,Mimura2005b,Florent2008}, and computer vision \cite{Tanaka2002}. 
These techniques play an important role in our modern life. 
\par
Equilibrium statistical mechanics is used to evaluate the performance of information processing problems 
that are formulated using statistical models with a large degree of freedom. 
These problems can be reformulated as a computation of marginal probabilities on factor graphs. 
Some iterative methods, which can be used to calculate marginal probabilities approximately, e.g., belief propagation, 
give us effective algorithms as iterative equations with parallel update. 
In general, the theoretical treatment of the dynamics of disordered systems is complicated compared with that of equilibrium statistical mechanics. 
It's important to understand the dynamical property of iterative algorithms such as a belief-propagation-based one in order to improve its performance. 
\par
In this study, we treat the continuous Hopfield model \cite{Hopfield1982} with the zero-temperature parallel dynamics as a typical example. 
There are many studies using the generating functional analysis (GFA) \cite{Coolen2000} of the Hopfield model with various dynamics. 
The GFA allows us to study the dynamics of an infinitely large disordered system in an exact way. 
For instance, Crisanti and Sompolinsky analyzed the asymmetric spherical-state Hopfield model with Langevin dynamics \cite{Crisanti1987}. 
Crisanti and Sompolinsky also analyzed an asymmetric Ising system with sequential-updated Glauber dynamics \cite{Crisanti1988}. 
Rieger analyzed the binary-state Hopfield model with sequential-updated Glauber dynamics \cite{Rieger1988}. 
Gardner {\it et al.} analyzed the Hopfield model with the zero temperature parallel dynamics \cite{Gardner1987}. 
Their approach can be used to calculate magnetization after several time steps but not to treat continuous states. 
\par
Kawamura {\it et al.} have found the sequential binary Hopfield model shows chaotic behaviors \cite{Kawamura2003, Kawamura2004}. 
In the case of the sequential binary Hopfield model with the Glauber dynamics, 
Kawamura and Okada have shown that the GFA result and the statistical neurodynamics (SND) \cite{Okada1995} result are identical \cite{Kawamura2002}. 
Boll$\rm\acute{e}$ {\it et al.} have investigated the relationship between the GFA and the SND using the binary Hopfield model with the Glauber dynamics \cite{Bolle2004}. 
Hashiguchi and Kawamura have studied the sequential continuous Hopfield model by using the SND \cite{Hashiguchi2006}. 
However, the relationship between the GFA and the SND in the continuous Hopfield model with parallel dynamics has remained unclear so far. 
Therefore, we here study this relationship. 
We apply the Langevin dynamics \cite{Crisanti1987} at zero temperature as the parallel dynamics to treat the continuous state. 
This dynamics has been applied to analysis of the batch minority games \cite{Galla2005}, etc. 
To confirm the validity of our analysis, we have performed some computer simulations for some typical conditions.

%~~~~~~~~~~~~~~~~~~~~~~~~~~~~~~~~~~~~~~~~~~~~~~~~~~~~~~~~~~~~~~~~~~~~~~~~~~~~~~~~~~~~~~~~~~~~

We consider a nonlinear system that consists of $N$ continuous variables $\sigma_i(t) \in \mathbb{R}$ at time $t$. 
In this paper, we treat the following updating rule: 
\begin{equation}
\sigma_i(t+1)=f(u_i(t)), \; u_i(t) \equiv h_i(\vec{\sigma}(t)) + \theta_i(t) \label{eq:deterministic_updating_rule}
\end{equation}
where $f:\mathbb{R}\to\mathbb{R}$ is an arbitrary function that called a {\it transfer function}. 
Here, $u_i(t)$ is called a {\it local field}, 
that includes the function $h_i(\vec{\sigma}(t))$ of $\vec{\sigma}(t) = {}^\dagger (\sigma_1(t),\cdots,\sigma_N(t))$ 
and a time-dependent external field $\theta_i(t)$ to calculate a response function. 
The operator ${}^\dagger$ denotes the transpose. 
We analyze the transient dynamics in the thermodynamical limit, i.e., $N \to \infty$. 
The Glauber dynamics can apply only the Ising system. 
Therefore, we first define the following deterministic updating rule for the variable $\sigma_i(t)$ at time $t$ as 
\begin{align}
\rho[\sigma_i(t+1)|\vec{\sigma}(t)] = \delta [ \sigma_i(t+1)-f(u_i(t)) ], \label{eq:updating_rule}
\end{align}
which can be regarded as the Langevin dynamics \cite{Crisanti1987} at zero temperature, 
where $\delta$ denotes the Dirac delta fuction. 
This dynamics represents Markovian dynamics, so the path probability $p[\vec{\sigma}(0),\cdots,\vec{\sigma}(t)]$ 
is simply given by the products of the individual transition probabilities $\rho[\vec{\sigma}(s+1)|\vec{\sigma}(s)]$ $= \prod_{i=1}^N \rho[\sigma_i(s+1)|\vec{\sigma}(s)]$ of the chain: 
\begin{equation}
p[\vec{\sigma}(0),\cdots,\vec{\sigma}(t)] = p[\vec{\sigma}(0)]\prod_{s=0}^{t-1}\rho[\vec{\sigma}(s+1)|\vec{\sigma}(s)]. \label{eq:def_path_probability}
\end{equation}
If we assume that we know the initial microscopic state precisely, it allows for a factorized distribution as $p[\vec{\sigma}(0)]=\prod_{i=1}^N p_i[\sigma_i(0)]$. 
\par
To analyze the transient dynamics of the continuous nonlinear system, we define a generating functional $Z[\vec{\psi}]$ as follows: 
\begin{align}
Z[\vec{\psi}]=
&\int_{\mathbb{R}^{(t+1)N}} \biggl( \prod_{s=0}^t d\vec{\sigma}(s) \biggr) p[\vec{\sigma}(0),\cdots,\vec{\sigma}(t)] \nonumber \\
&\qquad \times \exp \biggl(-i\sum_{s=0}^t\vec{\sigma}(s)\cdot\vec{\psi}(s) \biggr), 
\label{eq:def_Z}
\end{align}
where $\vec{\sigma}(s)={}^\dagger(\sigma_1(s),\cdots,\sigma_N(s))$ and $\vec{\psi}(s)={}^\dagger(\psi_1(s),\cdots,\psi_N(s))$. 
In a familiar way \cite{Coolen2000}, one can obtain from $Z[\vec{\psi}]$ all averages of interest by differentiation, 
i.e., spin-averages (or overlap), correlation-functions, and response-functions: 
\begin{align}
m_i(s)&=\langle \sigma_i(s) \rangle = i\lim_{\vec{\psi}\to\vec{0}}\frac{\partial \bar{Z}[\vec{\psi}]}{\partial \psi_i(s)}, \nonumber \\
C_{ij}(s,s')&=\langle \sigma_i(s)\sigma_j(s') \rangle = -\lim_{\vec{\psi}\to\vec{0}}\frac{\partial \bar{Z}[\vec{\psi}]}{\partial \psi_i(s) \partial \psi_j(s')}, \nonumber \\
G_{ij}(s,s')&=\frac{\partial \langle \sigma_i(s) \rangle}{\partial \theta_j(s')}=i\lim_{\vec{\psi}\to\vec{0}}\frac{\partial \bar{Z}[\vec{\psi}]}{\partial \psi_i(s) \partial \theta_j(s')}, 
\end{align}
where $\langle \; \rangle$ and $\overline{\cdots}$ denote the average with respect to the path probability $p[\vec{\sigma}(0),\cdots,\vec{\sigma}(t)]$ and the average over the disorder that is included in the generating functional $Z[\vec{\psi}]$, respectively.

%~~~~~~~~~~~~~~~~~~~~~~~~~~~~~~~~~~~~~~~~~~~~~~~~~~~~~~~~~~~~~~~~~~~~~~~~~~~~~~~~~~~~~~~~~~~~

In the continuous Hopfield model, the local field is defined as follows: 
\begin{align}
u_i(s) \equiv \! \! \sum_{j=1\;(\ne i)}^N \! \! \! \! J_{ij} \sigma_j(s) + \theta_i(s), \; J_{ij} = \frac 1N \sum_{\mu=1}^{\alpha N} \xi_i^\mu \xi_j^\mu, 
\end{align}
where $\alpha$ is the loading rate. 
The pattern components $\{\xi_i^\mu\}$ follow an identical and independent distribution 
\begin{align}
p(\xi_i^\mu) = \frac 12 \delta (\xi_i^\mu-1) + \frac 12 \delta (\xi_i^\mu+1). 
\end{align}
Performing an appropriate gauge transformation, we can put $\xi_i^1=1 \; (\forall i)$ without a loss of generality. 
We first separate the local field $u_i(s)$ by inserting the following $\delta$-function: 
$1=\int \prod_{s=0}^{t-1} \prod_{i=1}^N du_i(s) d\hat{u}_i(s)$ $\exp [i\hat{u}_i(s)\{u_i(s)-\sum_{j=1}^N J_{ij}\sigma_j(s) -\theta_i(s) \}]$.
Averaging the generating functional over the disorder $\{\xi_i^\mu\}$, we separate the relevant one-time and two-time order parameters: 
$m(s) = \frac 1N \sum_{i=1}^N \sigma_i(s)$,
$k(s) = \frac 1N \sum_{i=1}^N \hat{u}_i(s)$, 
$q(s,s') = \frac 1N \sum_{i=1}^N \sigma_i(s) \sigma_i(s')$, 
$Q(s,s') = \frac 1N \sum_{i=1}^N \hat{u}_i(s) \hat{u}_i(s')$, and 
$K(s,s') = \frac 1N \sum_{i=1}^N \sigma_i(s) \hat{u}_i(s')$. 
We then obtain the disorder-averaged generating functional 
\begin{align}
\bar{Z}[\vec{\psi}] = \int 
d\vec{m}d\hat{\vec{m}}
d\vec{k}d\hat{\vec{k}}
d\vec{q}d\hat{\vec{q}}
d\vec{Q}d\hat{\vec{Q}}
d\vec{K}d\hat{\vec{K}}
e^{N(\Psi+\Phi+\Omega)}, 
\end{align}
where 
\begin{align}
\Psi =& \; i\sum_{s=0}^{t-1}\{\hat{m}(s)m(s)+\hat{k}(s)k(s)-m(s)k(s)\} \nonumber \\
      & \; + i \sum_{s=0}^{t-1} \sum_{s'=0}^{t-1} \{ \hat{q}(s,s') q(s,s') + \hat{Q}(s,s') Q(s,s') \nonumber \\
      & \; \qquad \qquad \qquad \qquad + \hat{K}(s,s') K(s,s') \}, 
\end{align}
\begin{align}
\Phi =& \; \frac 1N \sum_{i=1}^N \ln \biggl\{ \int_{\mathbb{R}^{t+1}} \biggl( \prod_{s=0}^t d\sigma(s) \biggr) p[\sigma(0)] \int \delta h \delta \hat{h} \nonumber \\
      & \; \times \biggl( \prod_{s=0}^{t-1} \delta[ \sigma(s+1) - f(h(s)) ] \biggr) \nonumber \\ 
      & \; \times \exp \biggl[ -i \sum_{s=0}^{t-1} \sum_{s'=0}^{t-1} \{ \hat{q}(s,s') \sigma(s) \sigma(s') \nonumber \\
      & \; \qquad + \hat{Q}(s,s') \hat{h}(s) \hat{h}(s') + \hat{K}(s,s') \sigma(s) \hat{h}(s') \} \nonumber \\
      & \; + i \sum_{s=0}^{t-1} \hat{h}(s) \{ h(s) - \theta_i(s) - \hat{k}(s) \xi_i^1 \} \nonumber \\
      & \; - i \sum_{s=0}^{t-1} \sigma(s) \hat{m}(s) \xi_i^1 - i \sum_{s=0}^t \sigma(s) \phi_i(s) \biggr] \biggr\}, 
\end{align}
and 
\begin{align}
\Omega
     =& \; \frac 1N \ln \int \delta\vec{x} \delta\hat{\vec{x}} \delta\vec{y} \delta\hat{\vec{x}} \nonumber \\
      & \; \exp \biggl[ i \sum_{\mu=2}^{\alpha N} \sum_{s=0}^{t-1} \{ \hat{x}_\mu(s) x_\mu(s) + \hat{y}_\mu(s) y_\mu(s) \nonumber \\
      & \; \qquad \qquad \qquad \qquad - x_\mu(s) y_\mu(s) \} \nonumber \\
      & \; - \frac 12 \sum_{\mu=2}^{\alpha N} \sum_{s=0}^{t-1} \sum_{s'=0}^{t-1} \{ \hat{x}_\mu(s) Q(s,s') \hat{x}_\mu(s') \nonumber \\
      & \; \qquad + \hat{x}_\mu(s) K(s,s') \hat{y}_\mu(s') + \hat{y}_\mu(s) K(s,s') \hat{x}_\mu(s') \nonumber \\
      & \; \qquad + \hat{y}_\mu(s) q(s,s') \hat{y}_\mu(s') \biggr], 
\end{align}
with 
$\delta h = \prod_{s=0}^{t-1} \frac{dh(s)}{\sqrt{2\pi}}$, 
$\delta \hat{h} = \prod_{s=0}^{t-1} \frac{d\hat{h}(s)}{\sqrt{2\pi}}$, 
$\delta \vec{x} = \prod_{s=0}^{t-1} \prod_{\mu=2}^{\alpha N} \frac{dx_\mu(s)}{\sqrt{2\pi}}$, 
$\delta \hat{\vec{x}} = \prod_{s=0}^{t-1} \prod_{\mu=2}^{\alpha N} \frac{d\hat{x}_\mu(s)}{\sqrt{2\pi}}$, 
$\delta \vec{y} = \prod_{s=0}^{t-1} \prod_{\mu=2}^{\alpha N} \frac{dy_\mu(s)}{\sqrt{2\pi}}$, and 
$\delta \hat{\vec{y}} = \prod_{s=0}^{t-1} \prod_{\mu=2}^{\alpha N} \frac{d\hat{y}_\mu(s)}{\sqrt{2\pi}}$. 
\par
The disorder-averaged generating functional is for $N\to\infty$ dominated by a saddle-point. 
Applying the saddle-point method, we arrive at the following result. 
Some macroscopic variables are found to vanish at the relevant physical saddle-point. 
The remaining variables can all be expressed in terms of three macroscopic observables, namely, the overlaps $m(s)$, the single-site correlation functions $C(s,s')$ and the single-site response functions $G(s,s')$. 
We then arrive at the following saddle-point equations: 
\begin{eqnarray}
m(s)&=&\llangle \sigma(s) \rrangle, \label{eq:sp_m} \nonumber \\
C(s,s')&=&\llangle \sigma(s) \sigma(s') \rrangle, \label{eq:OrderParameters} \\
G(s,s')&=&
\left\{
\begin{array}{ll}
\alpha^{-1/2} \llangle \sigma(s)(\vec{R}^{-1}\vec{v})(s') \rrangle, & {\rm for} \; s>s' \\
0, & {\rm for} \; s \le s'.
\end{array}
\right. \nonumber
\end{eqnarray}
The average over the effective path measure is given by 
\begin{align}
\llangle g(\vec{\sigma},\vec{v}) \rrangle 
\equiv & \int_{\mathbb{R}^t} {\cal D}\vec{v} \int_{\mathbb{R}^{t+1}} \biggl( \prod_{s=0}^t d\sigma(s) \biggr) g(\vec{\sigma},\vec{v}) \nonumber \\
& \quad \times p[\sigma(0)] \prod_{s=0}^{t-1} \delta [ \sigma(s+1)-f(u(s)) ], \label{eq:FiniteTempPathAverage}
\end{align}
with 
\begin{align}
u(s) &= m(s)+\theta(s)+\sqrt{\alpha}v(s)+(\vec{\Gamma} \vec{\sigma})(s), \nonumber \\
\vec{R} &= (\vec{1}-\vec{G})^{-1} \vec{C} (\vec{1}-{}^\dagger \vec{G})^{-1}, \label{eq:SaddlePointEquations} \\
\vec{\Gamma} &= \alpha(\vec{1}-\vec{G})^{-1}\vec{G}, \nonumber
\end{align}
and ${\cal D}\vec{v} \equiv |2\pi \vec{R}|^{-1/2} d\vec{v}e^{-\frac 12 \vec{v}\cdot \vec{R}^{-1}\vec{v}}$. 
Here, $\vec{C}$, $\vec{G}$, $\vec{R}$ and $\vec{\Gamma}$ are matrices having matrix elements $C(s,s')$, $G(s,s')$, $R(s,s')$, and $\Gamma(s,s')$, respectively. 
$\vec{\sigma}$ is the vector ${}^\dagger (\sigma(0),\cdots,\sigma(t))$. 
The terms $(\vec{R}^{-1}\vec{v})(s)$ and $(\vec{\Gamma} \vec{\sigma})(s)$ denote the $s$th element of the vectors $\vec{R}^{-1} \vec{v}$ and $\vec{\Gamma} \vec{\sigma}$, respectively. 
The matrix $\vec{1}$ denotes a unit matrix. 
Equations (\ref{eq:OrderParameters})-(\ref{eq:SaddlePointEquations}) completely describe the dynamics of this continuous nonlinear system. 
In the limit where $t\to\infty$, the term $(\vec{\Gamma} \vec{\sigma})(s)$ in eq. (\ref{eq:SaddlePointEquations}) can be regarded as the Onsager reaction term in equilibrium statistical mechanics.

%~~~~~~~~~~~~~~~~~~~~~~~~~~~~~~~~~~~~~~~~~~~~~~~~~~~~~~~~~~~~~~~~~~~~~~~~~~~~~~~~~~~~~~~~~~~~

\begin{figure}[t]%[htbp]
\begin{center}
\includegraphics[width=.9\linewidth,keepaspectratio]{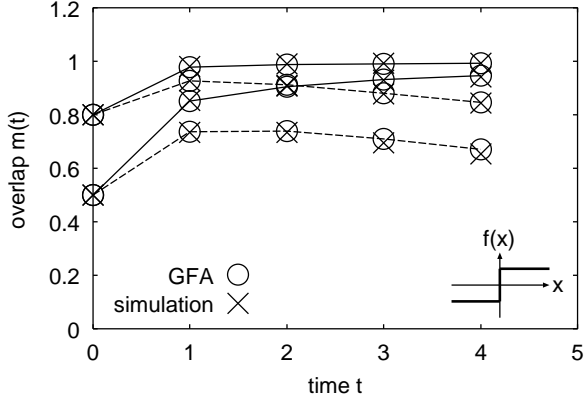}
\caption{
The first few time steps of the dynamics of the binary Hopfield model predicted using the GFA with $\alpha=0.12$ (circles and solid lines) and $\alpha=0.20$ (circles and dashed lines). 
The transfer function is $f(x)=\sgn(x)$. 
Computer simulations (crosses) were performed with $N=10,000$ from 10 experiments. 
}
\label{fig:Simulation1}
\end{center}
\end{figure}

\begin{figure}[t]%[htbp]
\begin{center}
\includegraphics[width=.9\linewidth,keepaspectratio]{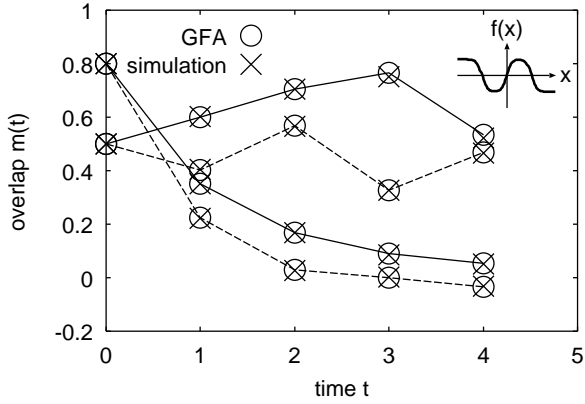}
\caption{
The first few time steps of the dynamics of the continuous Hopfield model predicted using the GFA with $\alpha=0.12$ (circles and solid lines) and $\alpha=0.20$ (circles and dashed lines). 
The transfer function is $f(x)=-\tanh[\gamma(x+k)]\tanh[\gamma x]\tanh[\gamma(x-k)]$ with $\gamma=5$ and $k=1$. 
Computer simulations (crosses) were performed with $N=10,000$ from 10 experiments. 
}
\label{fig:Simulation2}
\end{center}
\end{figure}

In most applications in the field of information theory, the initial state is usually given by a discrete distribution having the form 
\begin{align}
p[\sigma(0)]=\sum_{k=1}^K c_k \delta (\sigma(0)-\lambda_k), \label{eq:InitialProbability}
\end{align}
where $c_k \ge 0 \; (\forall k)$, $\sum_{k=1}^K c_k=1$ and $\lambda_k \in \mathbb{R} \;(\forall k)$. 
When the initial probability is given by the form of eq. (\ref{eq:InitialProbability}), 
the spin integral included in eq. (\ref{eq:FiniteTempPathAverage}) becomes a simple summation and the average over the effective path measure can be simplified to 
\begin{align}
& \llangle g(\vec{\sigma},\vec{v}) \rrangle = \int_{\mathbb{R}^t} {\cal D}\vec{v} \sum_{k=1}^K c_k \nonumber \\
& \qquad \times \biggl( \biggl. g(\vec{\sigma},\vec{v}) \biggr|_{\sigma(0)=\lambda_k, \sigma(1)=f(u(0)), \cdots, \sigma(t)=f(u(t-1))} \biggr). 
\label{eq:ZeroTempPathAverage}
\end{align}
The spin integral in eq. (\ref{eq:FiniteTempPathAverage}) can be performed in this system, 
therefore the effective single-spin dynamics $\sigma(s+1)=f(u(s))$ appears in eq. (\ref{eq:ZeroTempPathAverage}). 
The initial state is chosen stochastically from the initial distribution of eq. (\ref{eq:InitialProbability}), but dynamics is deterministic. 
The deterministic path of the initial state $\sigma(0)=\lambda_k$ appears with a probability $c_k$. 
The average over the effective path measure can be calculated from $K$ deterministic paths. 
\par
To validate the GFA, we performed computer simulations. 
We considered a case with the initial distribution $p[\sigma(0)]=\frac 12(1+m(0))\delta(\sigma(0)-1)+\frac 12(1-m(0))\delta(\sigma(0)+1)$ and $\theta(s)=0$. 
The first few time steps of the dynamics of the binary Hopfield model, which has the transfer function $f(x)=\sgn(x)$, predicted using the GFA with $\alpha=0.12$ and $\alpha=0.20$ are shown in Fig. \ref{fig:Simulation1}. 
The computer simulations were performed with $N=10,000$ from 10 experiments. 
Excellent agreement was found between the predictions of the GFA and the computer simulation results. 
The first few time steps of the dynamics of the continuous Hopfield model predicted using the GFA with $\alpha=0.12$ and $\alpha=0.20$ are shown in Fig. \ref{fig:Simulation2}. 
The transfer function is $f(x)=-\tanh[\gamma(x+k)]\tanh[\gamma x]\tanh[\gamma(x-k)]$ with $\gamma=5$ and $k=1$. 
Note that the parameter $\gamma$ is not the inverse temperature but an arbitrary parameter. 
In this case, computer simulations were also performed with $N=10,000$ from 10 experiments. 
When the transfer function $f$ is continuous, the GFA exhibited excellent consistency with the computer simulation results. 
\par
The relationship between the GFA and the SND is as follows. 
The SND analysis ignores the Onsager reaction term \cite{Okada1995}. 
In the binary-state parallel-update sequential processing Hopfield model, which does not have the Onsager reaction term, 
Kawamura and Okada showed that the GFA result was identical to the SND result \cite{Kawamura2002}. 
According to their discussion, if we neglect the Onsager reaction term, we can show that GFA result is equivalent to the SND result. 
When we neglect the Onsager reaction term in the GFA result of the continuous Hopfield model, which has the Onsager reaction term, the response function becomes 
\begin{align}
G(s,s')=
&\frac{\partial}{\partial \theta(s')} \int_{\mathbb{R}} {\cal D}v(s-1) \int_{\mathbb{R}} d\sigma(s) \; \sigma(s) \delta [\sigma(s) \nonumber \\
&- f(m(s-1)+\theta(s-1)+\sqrt{\alpha}v(s-1))], 
\end{align}
where ${\cal D}v(s-1)$ $\equiv$ $(2\pi R(s-1,s-1))^{-1/2} \times$ $e^{-\frac 12 v(s-1)^2/R(s-1,s-1)}$. 
This means that the response function becomes zero except for $G(s,s-1)$. 
Therefore, the $(s,s')$th element of the matrix $\vec{G}^n$ has the simple form $G^n(s,s')=\delta_{s,s'+n}\prod_{\tau=0}^{n-1} G(s-\tau,s-\tau-1)$. 
When the Onsager reaction term is neglected, eq. (\ref{eq:SaddlePointEquations}) can be represented in the recurrence relation form 
\begin{align}
R(s,s') = 
& C(s,s') \notag \\
& + G(s,s-1) G(s',s'-1) R(s-1,s'-1) \notag \\
& + \sum_{\lambda=0}^{s-1} C(\lambda,s') \sum_{\tau=\lambda+1}^s G(\tau,\tau-1) \notag \\
& + \sum_{\lambda=0}^{s'-1} C(s,\lambda) \sum_{\tau=\lambda+1}^{s'} G(\tau,\tau-1), \label{eq:recurrenceR}
\end{align}
by using the identity $(\vec{1}-\vec{G})^{-1}=\vec{1}+\sum_{n=1}^t \vec{G}^n$. 
The order parameters become 
\begin{align}
& m(s) = \int Dz \; f(\tilde{u}(s-1)), \notag \\
& C(s,s') = \int Du Dv Dz \; f(\bar{u}(s-1,s'-1)) \notag \\
& \qquad \qquad \qquad \qquad \; \; \; \times f(\bar{u}(s'-1,s-1)), \label{eq:OPs} \\
& G(s,s-1) = \frac1{\sqrt{\alpha R(s-1,s-1)}} \int Dz \; z f (\tilde{u}(s-1)), \notag 
\end{align}
where $\tilde{u}(s) \equiv m(s)$ $+ z\sqrt{\alpha R(s,s)}$ and $\bar{u}(s,s') \equiv m(s)$ $+ u\sqrt{\alpha[R(s,s)-R(s,s')]}$ $+ z \sqrt{\alpha R(s,s')}$.  
This result obtained using eqs. (\ref{eq:recurrenceR})-(\ref{eq:OPs}) coincides with the SND result \cite{Okada1995}. 
Namely, the SND result can be derived from the GFA result by neglecting the Onsager reaction term. 
\par
In summary, we applied the generating functional analysis to the continuous Hopfield model. 
The predictions of the GFA are in excellent agreement with computer simulation results for any conditions and transfer function. 
When the retarded self-interaction term is omitted, the GFA result becomes identical to the SND result. 
The GFA can also be applied to any continuous nonlinear system having the form of eq. (\ref{eq:updating_rule}) which includes some belief-propagation-based algorithms, e.g., multiuser detection problems in mobile communication systems.

%~~~~~~~~~~~~~~~~~~~~~~~~~~~~~~~~~~~~~~~~~~~~~~~~~~~~~~~~~~~~~~~~~~~~~~~~~~~~~~~~~~~~~~~~~~~~

This work was partially supported by a Grant-in-Aid for Scientific Research for Encouragement of Young Scientists (B) No. 18700230 from the Ministry of Education, Culture, Sports, Science and Technology of Japan.

\end{document}